\documentclass[twocolumn,amsmath]{revtex4-1}
\usepackage{hyperref,braket,amssymb,amsthm,graphicx,cleveref}
\newcommand{\C}{\mathcal{C}}
\newcommand{\prj}[1]{\ket{#1}\bra{#1}}
\newcommand{\idn}{\mathbb{I}}
\newcommand{\abs}[1]{\left\lvert{#1}\right\rvert}
\newcommand{\nrm}[1]{\left\lVert{#1}\right\rVert}
\DeclareMathOperator{\tr}{Tr}
\newtheorem{thm}{Theorem}

\begin{document}
\title{Verifying the quantumness of a channel with an untrusted device}
\author{Matthew F. Pusey}
\email{m@physics.org}
\affiliation{Perimeter Institute for Theoretical Physics, 31 Caroline Street North, Waterloo, ON N2L 2Y5, Canada}
\date{March 24, 2015}
\begin{abstract}
Suppose one wants to certify that a quantum channel is not entanglement-breaking. I consider all four combinations of trusted and untrusted devices at the input and output of the channel, finding that the most interesting is a trusted preparation device at the input and an untrusted measurement device at the output. This provides a time-like analogue of EPR-steering, which turns out to reduce to the problem of joint measurability, connecting these concepts in a different way to other recent work. I suggest a few applications of this connection, such as a resource theory of incompatibility. This perspective also sheds light on why the BB84 key distribution protocol can be secure even with an untrusted measuring device, leading to an uncertainty relation for arbitrary pairs of ensembles.
\end{abstract}
\maketitle

The crucial distinction between separable and entangled quantum states has its analogue in the notion of an entanglement-breaking channel \cite{eb1,eb2}. An entanglement-breaking channel $\C$ can be thought of as measuring a POVM $\{E_\lambda\}$ on the input system, transmitting the classical information $\lambda$, and then preparing an element of the set of states $\{\rho_\lambda\}$ on the output system, so that
\begin{equation}
\C(\rho) = \sum_\lambda \tr(E_\lambda \rho)\rho_\lambda .
\label{ebreakdef}
\end{equation}

Verifying the entanglement of a quantum state with various levels of trust is now a well-studied topic. In the bipartite case, there are three main possibilities.
Most commonly both parties are trusted, allowing state tomography and/or the use of entanglement witnesses. In the other extreme, neither party is trusted, so that one must resort to testing Bell inequalities \cite{bell,bellreview}. The intermediate case where one party is trusted is known as the EPR-steering scenario \cite{steer}, where one can test steering inequalities \cite{expcrit}.

Since a channel is entanglement-breaking if and only if it gives a separable state when applied to half of a maximally entangled state, the problem of verifying that a channel is not entanglement-breaking could be reduced to the above possibilities. However that requires the ability to prepare entangled states, and if any devices are untrusted it requires parties at space-like separation. Here I focus on the situation where there are just two parties, Alice who prepares quantum systems and inputs them entirely into the channel, and Bob who performs measurements solely on the output of the channel. Assuming that the channel is realisable and non-trivial, Bob's actions must be in the future of the Alice's, ruling out the possibility of space-like separation.

In this setup there are now four situations, one more than in the bipartite entanglement case due to the lack of Alice-Bob symmetry in the setup: untrusted Alice and trusted Bob is fundamentally different from trusted Alice and untrusted Bob. After a brief discussion of all four cases, I will focus on trusted Alice and untrusted Bob.

Perhaps unsurprisingly, this bears a close resemblance to the EPR-steering scenario. But it turns out to boil down to the well-known problem of joint measurability of POVMs. Steering and joint measurability have already been found to be closely connected: lack of joint measurability is necessary and sufficient for the measurements to enable violations of steering inequalities \cite{sjm1,sjm2}. Here the natural translation of the steering scenario from states to channels reveals a connection that is perhaps even closer: the translated concept of steering \emph{is} lack of joint measurability.

I outline a few translations enabled by this connection. For example, the resource theory of steering \cite{quant,resource} is easily adapted into a resource theory of measurement incompatibility. Building on \cite{monogamygame}, it also gives a more transparent and general understanding of why the BB84 key distribution protocol \cite{bb84} is secure with an untrusted measuring device.

The basic idea of this paper is implicit in the recently defined notion of ``channel steering'' \cite{chan}, being the special case when (in the notation of \cite{chan}) $B$'s output is trivial, just like channel steering reduces to standard EPR-steering when $C$'s input is trivial.
Hence all of the results of \cite{chan} apply here. Nevertheless I think this special case is interesting enough to merit specific investigation, and that this lends further support to the definition of channel steering chosen in \cite{chan}.

\section{The four scenarios}\label{fourscen}
The most familiar scenario is when both Alice and Bob are trusted. Then Alice can use an informationally complete set of preparations, and Bob an informationally complete set of measurements, to together reconstruct $\C$. This is known as process tomography \cite{nc}. One can then use standard tools from entanglement theory \cite{entangrev} to check whether $\C$ is entanglement-breaking (or equivalently \cite{eb2}, if the Choi isomorphic state \cite{choi} is separable).

Now consider the case that Alice has an untrusted preparation device. She chooses a classical label $x$ and then the device inputs an unknown state $\tilde\rho_x$ into the channel. Since Bob's devices are still trusted he can do state tomography to determine the resulting output states $\rho_x = \C(\tilde\rho_x)$. But whatever he sees could always have come from the entanglement-breaking channel
\begin{equation}
  \tilde \C(\rho) = \sum_x \rho_x \braket{x|\rho|x}
\end{equation}
provided Alice's device simply encodes $x$ in the orthonormal basis $\{\ket{x}\}$.

When compared to the standard steering scenario, the key difference is that there is no way to rule out Alice's choice $x$ from being transmitted to Bob, because the very thing we are testing is a channel from Alice to Bob. (Note that, by considering a more elaborate scenario, an analogy with steering can be obtained wherein an untrusted party acts before a trusted one \cite{temporal}.)

If we extend our distrust of Alice to encompass Bob, then we have strictly less information, namely just probabilities $p(b|x,y) = \tr(\tilde E_{b|y}\C(\tilde \rho_x))$ from inputting a setting $y$ and receiving an outcome $b$ from Bob's device that implements unknown POVMs $\{\tilde E_{b|y}\}$. Since we can simulate any $\C(\tilde \rho_x)$ with an entanglement-breaking channel, this can be extended to $p(b|x,y)$. In fact it is trivial to produce any $p(b|x,y)$ whatsoever: Alice encodes $x$ in the $\{\ket{x}\}$ basis, the channel is $\tilde\C(\rho) = \sum_x \prj{x} \rho \prj{x}$, and Bob measures the POVM $\{\sum_x p(b|x,y)\prj{x}\}$. (Again, with more background assumptions, a ``time-like'' analogue to Bell inequalities was obtained by Leggett and Garg \cite{lg}. Similarly to the issue here of Alice's setting being sent to Bob, the Leggett-Garg scenario allows ``signalling in time'' \cite{nsit} that weakens the analogy to the Bell scenario.)

Last but not least, the final possibility is a trusted Alice and an untrusted Bob. Using a informationally complete set of preparations $\{\rho_x\}$, we can reconstruct the POVM $E_{b|y} = \C^\dagger(\tilde E_{b|y})$ that Bob's POVMs $\tilde E_{b|y}$ induces on the input of $\C$. If $\C$ is an entanglement-breaking channel then its adjoint is, by \cref{ebreakdef}
\begin{equation}
  \C^\dagger(E) = \sum_\lambda E_\lambda \tr(E \rho_\lambda),
\end{equation}
so that
\begin{equation}
  E_{b|y} = \sum_\lambda E_\lambda p(b|y,\lambda), \label{jointmeas}
\end{equation}
where $p(b|y,\lambda) = \tr(\tilde E_{b|y}\rho_\lambda)$. Conversely any $E_{b|y}$ of the form \cref{jointmeas} can be achieved by the entanglement-breaking channel $\tilde C(\rho) = \sum_\lambda \tr(E_\lambda \rho) \prj{\lambda}$ provided Bob measures the POVM $\{\sum_\lambda p(b|y,\lambda) \prj{\lambda}\}$.

Readers familiar with the EPR-steering scenario \cite{steer,me,quant} will notice the resemblance of \cref{jointmeas} to the basic definition of a set of sub-normalized steered states $\{\sigma_{b|y}\}$ that are compatible with a separable state, known as an ``unsteerable assemblage'':
\begin{equation}
  \sigma_{b|y} = \sum_\lambda p(\lambda)\sigma_\lambda p(b|y,\lambda),
\end{equation}
where the quantum states $\sigma_\lambda$ are known as ``local hidden states''. Indeed if we absorb $p(\lambda)$ into $\tilde \sigma_\lambda = p(\lambda)\sigma_\lambda$ then we have
\begin{equation}
  \sigma_{b|y} = \sum_\lambda \tilde\sigma_\lambda p(b|y,\lambda),\label{unsteer}
\end{equation}
so that the only difference with \cref{jointmeas} is that $\sum_\lambda \tr(\tilde\sigma_\lambda) = 1$ whereas $\sum_\lambda E_\lambda = \idn$.


By analogy we might call $\{E_\lambda\}$ a ``local hidden POVM'' and the corresponding $\{E_{b|y}\}$ unsteerable (the latter being somewhat counter-intuitive terminology since Bob has to act \emph{after} Alice). But in fact there is no need for new terminology because \cref{jointmeas} is just the statement that the $\{E_{b|y}\}$ are jointly measurable (i.e. the fixed POVM $\{E_\lambda\}$ is sufficient to reproduce all of the measurements using the classical post-processing $\{p(b|y,\lambda)\}$).

\begin{table}
  \begin{tabular}{|c|c|c|c|}
    \hline
    Alice & Bob & General data & Classical data \\\hline
    Good & Good & Channel $\C$ & $\C(\cdot) = \sum_\lambda \tr(E_\lambda \cdot) \rho_\lambda$\\
    Good & Bad & Induced POVMs $\{E_{b|y}\}$ & $E_{b|y} = \sum_\lambda E_\lambda p(b|y,\lambda)$\\
    Bad & Good & Output states $\{\rho_x\}$ & All (but see \cite{temporal}) \\
    Bad & Bad & Probabilities $\{p(b|x,y)\}$ & All (but see \cite{lg}) \\\hline
  \end{tabular}
  \caption{Summary of the four possible combinations of trusted (``good'') and untrusted (``bad'') devices at Alice's input and Bob's output. The ``general data'' column lists what can be determined in the scenario, with ``classical data'' being the form compatible with an entanglement-breaking channel.}
  \label{sumtable}
\end{table}

The four scenarios are summarised in \cref{sumtable}. From now on the focus will be a trusted Alice and untrusted Bob. The discussion of this case above is summarised by
\begin{thm}
A channel from a trusted Alice to an untrusted Bob can be shown not to be entanglement-breaking if and only if the measurements Bob induces on the input to the channel are not jointly measurable.
\end{thm}

As noted in the introduction, this scenario is a special case of channel steering \cite{chan}, so that the above result is a special case of \cite{chan}'s Proposition 1. Similarly, Theorem 8 of \cite{chan} specialises to
\begin{thm}
  A channel $\C$ allows Bob to induce incompatible measurements at Alice's input if and only if the corresponding Choi state $\rho_{AB}$ allows Bob to steer Alice.
\end{thm}

In other words, a trusted Alice and untrusted Bob can certify that $\C$ is not entanglement-breaking in the present ``time-like'' scenario if and only if they could certify that $\rho_{AB}$ is entangled in the standard ``space-like'' scenario.

\section{Simple translations between EPR-steering and joint measurability}

In the usual EPR-steering scenario an assemblage may happen to satisfy $\sum_b \sigma_{b|y} = \idn/d$ (i.e. Alice's reduced state is maximally mixed). In this case we can simply rescale the assemblage to a set of POVMs $E_{b|y} = d\sigma_{b|y}$. Comparing \cref{jointmeas} with \cref{unsteer} we see that $\{E_{b|y}\}$ will be jointly measurable if and only if $\{\sigma_{b|y}\}$ is unsteerable. This connection allows for the translation of some results between scenarios, particularly existence results.

It was already noted in \cite{sjm2} that EPR-steering inequalities and joint measurability inequalities are closely related. The scaling argument makes this particularly transparent in the case of a linear EPR-steering inequality \cite{expcrit}, which is a set of Hermitian operators $F_{b|y}$ such that
\begin{equation}
  \sum_{b,y} \tr(F_{b|y}\sigma_{b|y}) \leq L
\end{equation}
for all unsteerable assemblages $\{\sigma_{b|y}\}$. Since a jointly measurable $\{E_{b|y}\}$ is $d$ times an unsteerable assemblage, we must have
\begin{equation}
  \sum_{b,y} \tr(F_{b|y}E_{b|y}) \leq dL
\end{equation}
for all jointly measurable $\{E_{b|y}\}$. Similarly, since an incompatible $\{E_{b|y}\}$ is $d$ times a steerable assemblage, and all steerable assemblages violate a linear EPR-steering inequality \cite{expcrit}, all incompatible $\{E_{b|y}\}$ will violate a linear joint measurability inequality obtained through this translation.

For example, the simple steering inequality given as Eq. (63) in \cite{expcrit} gives the joint measurability inequality
\begin{equation}
  \tr\left(X(E_{1|1} - E_{2|1}) + Y(E_{1|2} - E_{2|2})\right) \leq 2\sqrt{2} \label{jointexample}
\end{equation}
for two binary qubit POVMs $\{E_{b|1}\}$ and $\{E_{b|2}\}$, where $X$ and $Y$ are Pauli matrices.

Perhaps the most important fact about EPR-steering is that there are bipartite entangled states that nevertheless do not exhibit steering under projective measurements. The classic example \cite{steer}, based on a model by Werner \cite{werner}, involves maximally mixed reduced states. Hence they can be translated into examples of non-entanglement breaking channels where nevertheless projective measurements by Bob can only induce jointly measurable POVMs on the channel input. The simplest example would be the qubit channel
\begin{equation}
  \C_p(\rho) = p\rho + (1-p)\frac{\idn}2.
\end{equation}

Translating \cite{werner,steer} we see $\C_p$ is entanglement-breaking if and only if $p \leq \frac13$, whereas projective measurements on the output induce jointly measurable POVMs on the input whenever $p \leq \frac12$ (and the non-entanglement-breaking can be verified with untrusted Bob whenever $p > \frac12$). Very recently \cite{POVMlhs} it was noticed that a model by Barrett \cite{POVMlhv} can be adapted to show, in this scenario, that \emph{all} POVMs on the output induce jointly measurable POVMs on the input whenever $p \leq \frac5{12} \approx 0.41$.

As an example in the opposite direction, consider structures of joint measurability. Briefly, for a finite set of measurements $M = \{M_1, \dotsc, M_n\}$, the joint measurability structure is given by listing all of the (maximal) subsets of $M$ that are jointly measurable. The extreme cases are $\{M\}$, meaning that all the measurements can be measured together, and $\{ \{M_1\}, \dotsc, \{M_n\} \}$, meaning that every pair of measurements is incompatible. A recent result \cite{arbjoint} states that \emph{any} such structure can be realised in (finite-dimensional) quantum theory. Translating (i.e. dividing everything by $d$), we see that the corresponding structures of unsteerability in the EPR-steering scenario are also as rich as possible. For example, there is a three-settings assemblage that is steering even though the sub-assemblage consisting of any pair of settings is unsteerable - the EPR-steering equivalent of the so-called Specker's scenario \cite{speckerscenario,speckerscenario2} in which three measurements cannot all be jointly measured, even though any pair of them can.

However, the scenarios are not isomorphic, and so not every statement that is natural in one scenario will translate to a statement that is natural in the other. For example, any pair of two-outcomes measurements $\{E_{b|y}\}$ that are not jointly measurable can be used to violate the CHSH \cite{chsh} inequality by choosing an appropriate bipartite state and pair of measurements $\{E_{a|x}\}$ for the other party \cite{jointCHSH}. In the EPR-steering scenario one might therefore wonder if any steerable two-setting two-outcome assemblage $\{\sigma_{b|y}\}$ can be used to violate the CHSH inequality by choosing an appropriate $\{E_{a|x}\}$ to form the probabilities $\tr(E_{a|x}\sigma_{b|y})$, at least when $\sum_b \sigma_{b|y} = \idn/d$. In fact a direct translation of \cite{jointCHSH} doesn't establish this because there is nothing to play the role of the bipartite state (which in general does not have a maximally mixed reduced state and is therefore not isomorphic to a channel). An explicit example of a steerable two-setting two-outcome qubit assemblage (with $\sum_b \sigma_{b|y} = \idn/2$) that fails to violate the CHSH inequality is

\begin{align}
\sigma_{1|1} &= 0.3\prj{0},
\\\sigma_{2|1} &= 0.2\prj{0} + 0.5\prj{1},
\\\sigma_{1|2} &= 0.3\prj{+},
\\\sigma_{2|2} &= 0.2\prj{+} + 0.5\prj{-}.
\end{align}
Using semi-definite programming is easy to check that this assemblage is steering and yet the optimum $\{E_{a|x}\}$ give a value for CHSH's expression of approximately $1.7$,  which satisfies the local upper bound of $2$. Another way of stating this fact is that ``choosing an appropriate bipartite state'' is essential to the result of \cite{jointCHSH}, one cannot simply use a maximally entangled state for any pair of measurements.

\section{The resource theory of measurement incompatibility}
Most of our quantitative understanding of entanglement \cite{entangrev} comes from thinking of it as a resource theory: allowing local operations and classical communication for free, how `useful' is a given multipartite quantum state (the basic objects of the theory)? Recently, to understand how best to quantify steering \cite{quant,quant2}, a similar ``resource theory of steering'' \cite{resource} has been developed.

In the resource theory of steering, the free operations are local quantum operations by the trusted party, local classical operations on the input and output of the untrusted party, and one-way classical communication from the trusted to the untrusted party.

\begin{figure*}
\begin{center}
\includegraphics{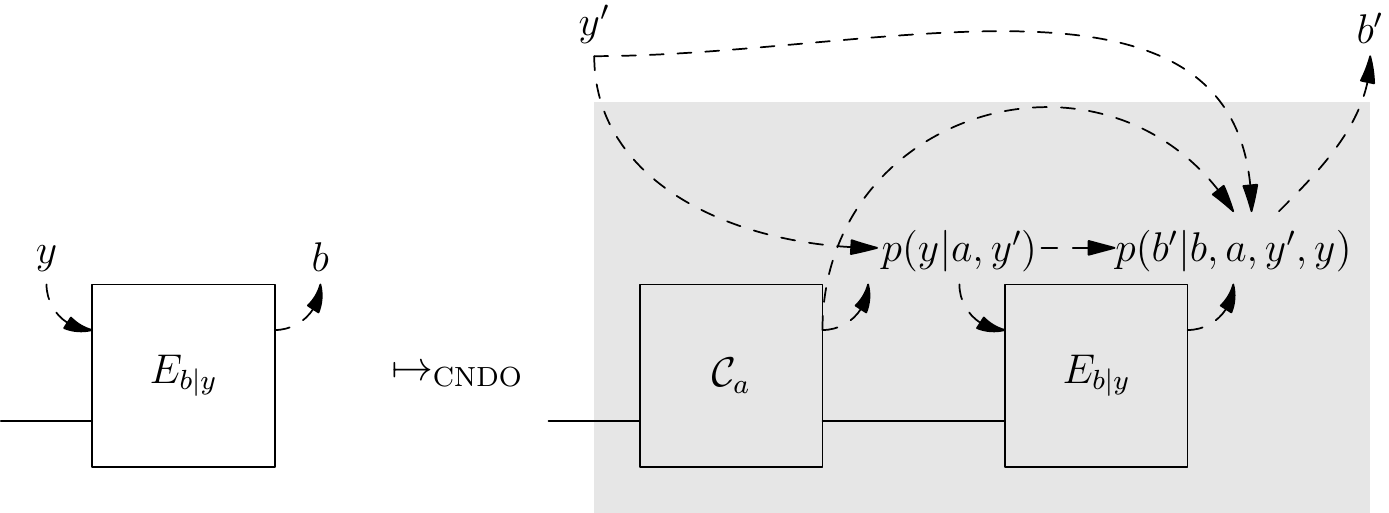}
\end{center}
\caption{A compatibility non-decreasing operation (CNDO) transforms a set of POVMs $\{ E_{b|y} \}$ into a set $\{ F_{b'|y'} \}$ enclosed in the grey box (cf.\ figure 1 of \cite{resource}). Inside the box there is a fixed quantum instrument $\{\C_a\}$ and classical conditional probability distributions $p(y|a,y')$ and $p(b'|b,a,y',y)$. Solid lines carry quantum systems, dashed lines indicate classical information.}
\label{CNDOfig}
\end{figure*}

This translates naturally into a resource theory of measurement incompatibility. The basic objects are sets of POVMs $\{E_{b|y}\}$. The free operations, which might be called \emph{compatibility non-decreasing operations} (CNDO) consist of prepending these POVMs with a fixed quantum instrument $\{\C_a\}$, classically processing the new input $y'$ along with $a$ into a new choice $y$ from amongst the original POVMs, and classically processing outcome $b$, along with $(a,y',y)$, into a new output $b'$, as in \cref{CNDOfig}. The requirement that the initial quantum instrument is fixed (i.e. independent of $y'$), is equivalent to the restriction that the untrusted party cannot send messages to the trusted party in the resource theory of steering.

What are the ``free resources'' in this theory, analogous to the separable states in entanglement theory? We will now see that they are simply jointly measurable POVMs. We can turn an arbitrary input set of POVMs $\{E_{b|y}\}$ into an arbitrary set of jointly measurable POVMs $\{ F_{b|y} \}$ by prepending the joint measurement, feeding an arbitrary state into $\{E_{b|y}\}$ and ignoring the output, and then classically processing the output of the joint measurement to obtain the appropriate output $b$ for a given input $y$. On the other hand, if we start with a jointly measurable POVM $\{ E_{b|y} \}$, the CNDO can only give us another jointly measurable POVM $\{ F_{b|y} \}$ because prepending the joint measurement of $\{E_{b|y}\}$ with the fixed quantum instrument gives us a joint measurement of $\{ F_{b|y} \}$.

Incompatible POVMs are then a resource: for example, for any such POVMs there exist entangled states for which the POVMs would demonstrate EPR-steering \cite{sjm1,sjm2}. In the scenario considered above, any incompatible POVMs can be used by an untrusted party to show that they are receiving a non-entanglement-breaking channel from a trusted party, at least in the case of a unitary channel (whose adjoints, which are also unitary channels, always map incompatible POVMs to incompatible POVMs).

The main results of \cite{resource} translate into results about measurement incompatibility. For example, in analogy with the ``steerable weight'' we can define the ``incompatible weight'' of a set of POVMs $\{E_{b|y}\}$ as the smallest $\nu \geq 0$ such that
\begin{equation}
E_{b|y} = \nu F_{b|y} + (1-\nu) G_{b|y}
\end{equation}
for arbitrary POVMs $\{F_{b|y}\}$ and jointly measurable POVMs $\{ G_{b|y} \}$. Evidently the incompatible weight is zero if and only if the $\{E_{b|y}\}$ are jointly measurable, and an argument similar to Appendix D of \cite{resource} shows that it is non-increasing under CNDO and can therefore be called an ``incompatibility monotone'' \footnote{A definition of ``incompatibility monotone'' has also been given in \cite{noisemeas}, but with a smaller set of free operations: the only classical processing allowed there is the complete swapping of two POVMs.}\nocite{noisemeas}.

As another example, in \cref{binaryconvertproof} it is argued that Theorem 4 of \cite{resource} gives
\begin{thm}\label{binaryconvert}
Suppose $\{E_{b|y}\}$ and $\{F_{b|y}\}$ are both pairs of distinct non-trivial two-outcome projective measurements on a qubit. Then the first can be converted into the second by CNDO if and only if it can be converted by simply prepending a unitary. \footnote{Notice that all pairs of distinct non-trivial two-outcome projective measurements on a qubit are always incompatible, and that non-distinct pairs, or pairs containing the trivial projective measurement $\{\idn\}$, are always jointly measurable. Recalling that any jointly measurable POVMs can be produced using CNDO, and that jointly measurable POVMs cannot be turned into incompatible POVMs, we see that \cref{binaryconvert} covers the only non-trivial case concerning pairs of two-outcome projective measurements on a qubit.}
\end{thm}
Compare the pair of measurements defined by the eigenstates of $X$ and $Z$ with the pair defined by $X$ and $\sqrt{1-\epsilon}X + \sqrt{\epsilon}Z$ for small $\epsilon>0$. Since the second pair would be jointy measurable for $\epsilon=0$, one might think that the first pair is ``more incompatible'' than the second. But since the above theorem tells us that neither pair can be converted to the other by CNDO, there not only exists an incompatibility monotone that assigns a higher value to the first than the second, but also another monotone that does vice versa. \footnote{The existence of both monotones follows from the fact that, in any resource theory, one resource is convertible to another if and only if every monotone assigns a greater or equal value to the first than the second. \cite{genresource}.}\nocite{genresource}

\section{Quantum Key Distribution}
An entanglement-breaking channel as in \cref{ebreakdef} can trivially be extended to give additional parties the same access to Alice's input as Bob has, for example
\begin{equation}
 \C'(\rho) = \sum_\lambda \tr(E_\lambda \rho) \rho_\lambda \otimes \rho_\lambda.
\end{equation}
However, channels which are not entanglement-breaking have limits on such extendibility, giving a channel analogue to the monogamy of entanglement \cite{entangrev}. Hence a necessary condition for the security of a ``prepare and measure'' quantum key distribution protocol is that Alice and Bob verify that the channel between them is not entanglement-breaking.

From the analysis of \cref{fourscen} we therefore see that prepare-and-measure key distribution is impossible with completely untrusted preparation devices (although rather weak forms of trust can suffice \cite{semidev}). But since the non-entanglement-breaking property \emph{can} be verified with an untrusted measuring device, it is plausible that quantum key distribution is possible in this scenario.

\subsection{BB84}

For example, the parties in the BB84 protocol \cite{bb84} are doing everything needed to verify a violation of the joint measurability inequality \cref{jointexample} (with $Z$ in place of $Y$). And indeed security proofs for the BB84 protocol have been provided for this scenario \cite{direct1,direct2,direct3,arbindiv,uncertainsmooth,oneside,monogamygame,cloning,cloning2}, with the distrust of Bob's devices emphasized particularly in \cite{uncertainsmooth,oneside,monogamygame}. This type of scheme was dubbed ``one-sided device independent quantum key distribution'' in \cite{oneside}, and entanglement-based versions were shown to rest upon EPR-steering.

The analyses \cite{uncertainsmooth,oneside,monogamygame} are all based on considering equivalent entanglement-based protocols, following \cite{wobell}. A more direct approach, along the lines pursued in earlier works \cite{direct1,direct2,direct3} and more recently in \cite{cloning,cloning2}, may add conceptual clarity, as well as enabling consideration of cases where the average state prepared by Alice depends on her choice of basis, which would violate the no-signalling principle in the entanglement-based scenario but is perfectly possible in an experiment where the states are prepared directly. (Although such cases can also be addressed using different entangled states for each of Alice's bases, as in, for example, \cite{arbindiv}.)

The analyses of \cite{monogamygame} can fairly straightforwardly be adapted to apply directly to the ``prepare and measure scenario'' we are considering. They are based on ``monogamy of entanglement games'', which are defined by $n$ POVMs $\{E_{x|\theta}\}$ for Alice. Bob and Charlie then choose a state $\rho_{ABC}$ and POVMs $\{ F_{x|\theta}\}$ and $\{ G_{x|\theta}\}$ with the aim of maximizing quantities such as
\begin{equation}
  p_\text{win} = \frac1n \sum_{\theta,x} \tr\left( (E_{x|\theta} \otimes F_{x|\theta} \otimes G_{x|\theta})\rho_{ABC} \right).\label{theirwin}
\end{equation}

When considering key distribution, the eavesdropper would choose $\rho_{ABC}$ and have full control of Charlie.

Translating to the present scenario, a game would be defined by $n$ ensembles $\{\rho_{x|\theta}\}$ for Alice. Bob and Charlie would then choose a channel $\C$ from $A$ to $BC$ and POVMs as before with the aim of maximizing quantities such as
\begin{equation}
  p_\text{win} = \frac1n \sum_{\theta,x} \tr\left( (F_{x|\theta} \otimes G_{x|\theta})\C(\rho_{x|\theta}) \right).\label{mywin}
\end{equation}

For an intuitive idea of the connection between such a game and key distribution, suppose a game is such that $p_\text{win}$ is bounded away from $1$. Then if, whichever basis $\theta$ is announced by Alice, Bob's POVM $F_{x|\theta}$ is able to recover Alice's bit $x$ perfectly, it must be that Charlie's POVM cannot also recover it perfectly. In other words, the strength of the correlation between Alice and Bob's copy of $x$ provide limits anybody else's access to $x$. The assumption that $\C$ is independent of $\theta$ must be enforced by Alice keeping $\theta$ secret until Bob's output system is in his custody. Since the BB84 protocol is based on Alice preparing random eigenstates of $Z$ and $X$, it will be related to the ``BB84 game'' $\rho_{0|0} = \prj{0}/2$, $\rho_{1|0} = \prj{1}/2$, $\rho_{0|1} = \prj{+}/2$, $\rho_{1|1} = \prj{-}/2$.

With minor modifications, outlined in \cref{monogamytranslate}, the main results of \cite{monogamygame} can be translated to this type of game. In particular, this should facilitate the study of prepare-and-measure protocols that cannot be converted into entanglement-based protocols because of the no-signalling issue discussed above. It also provides an interesting type of uncertainty relation. Fix some pair of ensembles $\{\rho_{x|0}\}$ and $\{\rho_{x|1}\}$ on a $d$-dimensional Hilbert space $A$. For a given channel $\C$ from $A$ to $BC$, we can define $p_\text{guess}(X|B\Theta)$ to be the optimum probability of guessing $x$ given $\tr_C(\C(\rho_{x|\theta}))$, averaged over the uniformly random choice $\theta$ of ensemble. Similarly $p_\text{guess}(X|C\Theta)$ denotes the optimum probability of guessing $x$ given $\tr_B(\C(\rho_{x|\theta}))$. A result of \cite{monogamygame} can then be translated into the statement that for all $\C$,
\begin{equation}
  p_\text{guess}(X|B\Theta) + p_\text{guess}(X|C\Theta) \leq d(m + \sqrt{c}) \label{uncerteqn},
\end{equation}
where the two ensembles $\{\rho_{x|\theta}\}$ are characterized by $c = \max_{x,z}\nrm{\sqrt{\rho_{x|0}}\sqrt{\rho_{z|1}}}^2$ and $m = \max_{x,\theta}\tr(\rho_{x|\theta})$.
 Note that rather than the usual two incompatible measurements on some state, this uncertainty relation is about two ensembles $\{\rho_{x|0}\}$ and $\{\rho_{x|1}\}$ on $A$ being fed into some channel to $BC$. Such uncertainty relations have been considered before, for example Corollary 6 of \cite{chanuncert1}, and \cite{chanuncert2}. However those works use an equivalence to the usual measurement uncertainty scenario that only works if $\sum_x \rho_{x|0} = \sum_x \rho_{x|1}$. By contrast, \cite{cloning} provided a similar type of uncertainty relation (called there a ``cloning bound'') without this limitation. \Cref{uncerteqn} shares this feature.

\subsection{B92}

Another interesting prepare-and-measure key distribution protocol was proposed by Bennett \cite{b92}. It is in some sense ``minimal'' in that Alice chooses between preparing just two (without loss of generality, qubit) states $\ket{\psi_1}$ and $\ket{\psi_2}$, which can be arbitrary non-orthogonal pure states. Bob is supposed to choose between measuring either $\{ E_{1|1}, E_{2|1} \} = \{\idn - \prj{\psi_2}, \prj{\psi_2}\}$ or $\{ E_{1|2}, E_{2|2} \} = \{ \idn - \prj{\psi_1}, \prj{\psi_1} \}$.

Suppose that Bob's measuring devices are untrusted. Let us examine what we can learn about the measurements $E_{b|y}$ he induces on Alice's input. If we find that for $i \neq j$, $\braket{\psi_j|E_{1|i}|\psi_j} = 0$, then $E_{1|i} \leq \idn - \prj{\psi_j}$. If this is all we verify then the eavesdropper may be unambiguously discriminating \cite{usd1,usd2} between the $\ket{\psi_i}$, and then giving Bob's measuring device classical instructions to output 1 for the $E_{1|i}$ measurement if the eavesdropper has determined that the state is not $\ket{\psi_i}$, and $2$ otherwise. This is similar to the standard attack on Bennett's protocol in the presence of losses, except that the eavesdropper's increased control over the measuring device means that in place of the losses the eavesdropper can simply instruct Bob's device to give result $2$ for both measurements.

However, if we further verify that $\braket{\psi_i|E_{1|i}|\psi_i} = 1 - \abs{\braket{\psi_1|\psi_2}}^2$ (a mild extension to the protocol in \cite{b92}, which suggests throwing away some of the data needed to estimate this probability), then in fact Bob must be inducing the correct measurements. Since these measurements are incompatible, the parties have certified that the channel is not entanglement-breaking and there is hope for the protocol to be secure. A full security proof would of course require moving beyond the i.i.d.\ assumptions implicit in this discussion.

\section{Conclusions}
When translating EPR-steerability from bipartite states to channels the problem of joint measurability naturally emerges. I have outlined a few applications of this connection, in particular to define a resource theory of measurement incompatibility, but since the literatures on EPR-steering and joint measurability are both large there are bound to be many more.

The security of BB84 with an untrusted measurement device is clarified by this perspective, in particular the definitions and results of \cite{monogamygame} can be adapted to apply directly to the prepare-and-measure scenario. This also provides an uncertainty relation for ensembles. It may be fruitful to see if more powerful measurement uncertainty relations, such as those of \cite{uncertainsmooth}, can also be adapted to arbitrary pairs of ensembles. Finally, it would be interesting to further investigate whether the conceptually important B92 protocol can also be secure with an untrusted measurement device.

\begin{acknowledgments}
I thank Patrick Coles, Marco Piani and Erik Woodhead for a useful discussions and comments. Research at Perimeter Institute is supported in part by the Government of Canada through NSERC and by the Province of Ontario through MRI.
\end{acknowledgments}

\bibliography{steermeas}

\appendix

\section{Proof of \cref{binaryconvert}}\label{binaryconvertproof}
The ``if'' direction is trivial. For the ``only if'' direction, direction, suppose $\{E_{b|y}\}$ can be converted into $\{F_{b|y}\}$ by CNDO. That is, there exists a quantum instrument $\{\C_a\}$ and probability distributions $\{p(y|y',a)\}$, $\{p(b'|b,a,y',y)\}$ such that
\begin{equation}
F_{b'|y'} = \sum_{a,y,b} \C_a^\dagger(E_{b|y})p(y|y',a)p(b'|b,a,y',y)\label{CNDOeq}
\end{equation}

Since $\{F_{b'|y'}\}$ and $\{E_{b|y}\}$ are rank-one projectors, they can also be viewed as normalised quantum states. Hence the only difference between \cref{CNDOeq} and the corresponding EPR-steering problem is that $\sum_a \C_a^\dagger$ is not necessarily trace-preserving (but must be unital). However, the only properties of the corresponding maps used in the proof in Appendix D of \cite{resource} is complete positivity (in the form of the existence of a Kraus decomposition) and so the proof applies here as well.

\section{Translation guide for \cite{monogamygame}}\label{monogamytranslate}
Recall that we are interested in translating analyses from the form of \cref{theirwin} to the form of \cref{mywin}. We can bridge the gap somewhat by using the positive operator $\rho_{ABC}$ Choi isomorphic to $\C$, so that $\C(\rho_{x|\theta}) = \tr_A(\rho_{ABC}\rho_{x|\theta}^T)$. Then \eqref{mywin} becomes

\begin{equation}
  p_\text{win} = \frac1n \sum_{\theta,x} \tr\left( (\rho_{x|\theta}^T \otimes F_{x|\theta} \otimes G_{x|\theta})\rho_{ABC}) \right).
\end{equation}
which looks more like \cref{theirwin}. The two main differences to keep in mind are that $\sum_x \rho_{x|\theta}^T$ is some unit trace operator, which may depend on $\theta$, rather than $\idn$, and $\rho_{ABC}$ has trace $d$ (the dimension of Alice's Hilbert space) rather than trace 1.

With one exception, we can ignore the requirement on Choi states that $\tr_{BC}(\rho_{ABC}) = \idn$ because statements about all $\rho_{ABC}$ will in particular be true about ones satisfying the constraint. It would be interesting to study whether using this constraint could lead to tighter bounds.

In \cite{monogamygame} frequent use is made of the ``purification'' of Bob and Charlie's strategy. In our scenario that corresponds to dilating $\C$ into an isometry (giving the extra system to say, Charlie, who ignores it) and then purifying Bob and Charlie's measurements as before.

I will now outline the results of these techniques on the statements and proofs of the main theorems in \cite{monogamygame}.

The statement of \textbf{theorem 3} of \cite{monogamygame}, giving the optimum winning probability for the ``BB84 game'', survives unscathed, except that the game is rescaled as $\rho_{x|\theta} = \rho_{x|\theta}^T = F_x^\theta / 2^n$. The strategy used to show that the probability can be achieved does not translate into a valid channel (this is the ``one exception'' mentioned above). For $n=1$ consider instead the channel that measures $\{\prj{\phi}, \idn-\prj{\phi}\}$ and broadcasts the result $0$ or $1$ to Bob and Charlie to use as their guesses. It is easy to see that this gives the same probability of success and for $n>1$ it can be repeated $n$ times. The proof of optimality is essentially unchanged, in Eq. (7) of \cite{monogamygame} we pick up a factor of $2^n$ from the larger trace of $\rho_{ABC}$ but it cancels with the rescaling of $\rho_{x|\theta}$.

\textbf{Theorem 4} of \cite{monogamygame}, which bounds the winning probability of arbitrary games, can be converted into
\begin{equation}
  p_\text{win}(\mathsf{G}^{\times n}; \mathcal{Q}) \leq \abs{\mathcal{Q}}d^n\left( \frac{m(\mathsf{G})}{\abs{\Theta}} + \frac{\abs{\Theta} - 1}{\abs{\Theta}}\sqrt{c(\mathsf{G})} \right)^n, \label{mythm4}
\end{equation} 
where in addition to the usual maximal overlap
\begin{equation}
  c(\mathsf{G}) = \max_{\substack{\theta,\theta' \in \Theta \\ \theta \neq \theta'}} \max_{x,x' \in \mathcal{X}} \nrm{\sqrt{\rho_{x|\theta}}\sqrt{\rho_{x'|\theta'}}}^2,\label{cdef}
\end{equation}
we need the maximal probability
\begin{equation}
  m(\mathsf{G}) = \max_{\theta \in \Theta} \max_{x \in \mathcal{X}} \tr(\rho_{x|\theta}).\label{mdef}
\end{equation}
The factor of $d^n$ comes from the trace of $\rho_{ABC}$. In the proof's definition of $A$ and $B$ we can multiply $\idn_{\mathcal{T}^c}$ by $m(\mathsf{G})^{\abs{\mathcal{T}^c}} = m(\mathsf{G})^{n-t}$ and still have $B \geq A_q^\theta$ and $C \geq A_q^{\theta'}$ as required. This factor carries through to $\nrm{\sqrt{B}\sqrt{C}} \leq m(\mathsf{G})^{n-t} (\sqrt{c(\mathsf{G})})^t$ and ultimately into \eqref{mythm4}. The transposes on $\rho_{x|\theta}$ affect neither the norms nor traces in \cref{cdef,mdef} respectively.

\textbf{Theorem 5} of \cite{monogamygame}, which concerns the security of an entangled-based variant of BB84 with an arbitrary measurement device for Bob, translates into a statement of the security of the original BB84 protocol with an arbitrary measurement device. The state $\rho_{\Theta TXYE}$ is defined by Alice choosing a uniformly random $\theta$ and preparing the $n$-qubit BB84 ensembles $\rho_{x|\theta} = \prj{x^\theta}/2^n$ with a classical record of the $\theta$ and $x$ stored in Alice's registers $\Theta$ and $X$, inputting the $n$ qubits into an unknown channel to Bob and Eve, and then Alice announcing $\theta$ to Bob so that he can perform the corresponding unknown POVM $\{P_{x|\theta}\}$ and store the result in $Y$. Since $m(\mathsf{G}^{\times n}_\text{BB84}) = 2^{-n}$, \cref{mythm4} gives $p_\text{win}(\mathsf{G}^{\times n}_\text{BB84},\mathcal{Q}^n_{\gamma+\epsilon,0}) \leq \beta^n$ exactly as in \cite{monogamygame}. The remainder of the proof is unaltered.

Finally, \textbf{Theorem 8} of \cite{monogamygame}, which provides an uncertainty relation, becomes a statement about
\begin{equation}
  \rho_{XBC\Theta} = \sum_{x,\theta} \frac12 \prj{x}_X \otimes \C(\rho_{x|\theta}) \otimes \prj{\theta}_\Theta
\end{equation}
for a channel $\C$ from $A$ to $BC$ and a pair of ensembles $\{\rho_{x|0}\}$ and $\{\rho_{x|1}\}$ on the $d$-dimensional system $A$. The statement is that
\begin{equation}
  p_\text{guess}(X|B\Theta) + p_\text{guess}(X|C\Theta) \leq d(m + \sqrt{c}),
\end{equation}
and
\begin{equation}
  H_{\min}(X|B\Theta) + H_{\min}(X|C\Theta) \geq -2\log\left( d\frac{m + \sqrt{c}}2 \right),
\end{equation}
where $c = \max_{x,z}\nrm{\sqrt{\rho_{x|0}}\sqrt{\rho_{z|1}}}^2$ and $m = \max_{x,\theta}\tr(\rho_{x|\theta})$. The $d$ is from the normalization of the Choi state $\rho_{ABC}$, and the $m$ is from strengthening $\nrm{A_i^\theta} \leq 1$ to $\nrm{A_i^\theta} \leq m$.
\end{document}